\PassOptionsToPackage{dvipsnames}{xcolor} 
\PassOptionsToPackage{hyperfootnotes=false}{hyperref} 
\documentclass[acmsmall, screen, nonacm]{acmart}

\AtBeginDocument{%
  }

\usepackage{algorithm}

\usepackage{xcolor}

\newsavebox{\nrboxcontent}
\definecolor{findingboxbg}{HTML}{F2F2F2}
\definecolor{findingboxborder}{HTML}{404040}
\newenvironment{codebox}{%
  \setlength{\fboxsep}{4pt}%
  \setlength{\fboxrule}{0.4pt}%
  \begin{lrbox}{\nrboxcontent}
  \begin{minipage}{\dimexpr\linewidth-8pt-0.8pt\relax}
  \hspace*{4pt}\begin{minipage}[t]{\dimexpr\linewidth-8pt-0.8pt-2pt\relax}
}{%
  \end{minipage}
  \end{minipage}
  \end{lrbox}%
  \noindent\fcolorbox{gray!20}{gray!5}{\usebox{\nrboxcontent}}%
}
\newenvironment{findingbox}{%
  \par\vspace{6pt}%
  \setlength{\fboxsep}{2mm}%
  \setlength{\fboxrule}{1pt}%
  \begin{lrbox}{\nrboxcontent}
  \begin{minipage}{\dimexpr\linewidth-4mm-1pt\relax}
}{%
  \end{minipage}
  \end{lrbox}%
  \noindent\fcolorbox{findingboxborder}{findingboxbg}{\usebox{\nrboxcontent}}%
  \par\vspace{6pt}%
}

\makeatletter
\AtBeginEnvironment{findingbox}{\let\@vspace\@vspace@orig\let\@vspacer\@vspacer@orig}
\makeatother
\newenvironment{promptbox}[1]{%
  \setlength{\fboxsep}{5pt}%
  \setlength{\fboxrule}{0.4pt}%
  \begin{lrbox}{\nrboxcontent}
  \begin{minipage}{\dimexpr\linewidth-10pt-0.8pt\relax}
  \noindent\colorbox{black!15}{\parbox{\dimexpr\linewidth-10pt-0.8pt\relax}{\vspace{2pt}\noindent\small #1\vspace{2pt}}}\par\vspace{3pt}
  \scriptsize\ttfamily
}{%
  \end{minipage}
  \end{lrbox}%
  \noindent\fcolorbox{black!15}{gray!5}{\usebox{\nrboxcontent}}%
}
\usepackage{multirow}
\usepackage{makecell}

\usepackage{graphicx}
\usepackage{xspace}
\usepackage{enumitem}
\setlist[itemize]{leftmargin=1.2em}
\usepackage{listings}
\usepackage[utf8]{inputenc}
\usepackage[T1]{fontenc}
\definecolor{KWColor}{rgb}{0.37,0.08,0.25}
\definecolor{CommentColor}{rgb}{0.12,0.38,0.18}
\definecolor{StringColor}{rgb}{0.06,0.10,0.98}
\definecolor{darkred}{rgb}{0.75,0,0}
\definecolor{lightgrey}{rgb}{0.8,0.8,0.8}
\lstdefinestyle{Eclipse}{
  xleftmargin=0pt,
  basicstyle=\ttfamily\scriptsize,
  commentstyle=\color{CommentColor}\ttfamily\footnotesize,
  stringstyle=\color{StringColor},
  keywordstyle=\color{KWColor}\bfseries,
  escapeinside={/*@}{@*/}
}
\definecolor{diffaddbg}{RGB}{212,237,218}   
\definecolor{diffremovebg}{RGB}{220,90,90}

\usepackage{siunitx} 
\usepackage{setspace}
\usepackage{subcaption}
\usepackage{url}
\usepackage{cleveref}

\crefname{figure}{Figure}{Figures}
\crefname{table}{Table}{Tables}
\crefname{section}{Section}{Sections}
\crefname{equation}{Equation}{Equations}
\crefname{algorithm}{Algorithm}{Algorithms}
\crefname{lstlisting}{Listing}{Listings}

\newcommand{\toolname}{\textsc{NullRepair}\xspace}

\begin{document}

\newcommand{\theTypeInferenceFramework}{the Type Inference Framework\xspace}
\newcommand{\TheTypeInferenceFramework}{The Type Inference Framework\xspace}
\newcommand{\TIF}{TIF\xspace}

\newcommand{\conferencePageLimit}{20}

\newcommand{\todo}[1]{{\color{red}\bfseries [[#1]]}}

\newcommand{\manu}[1]{\todo{#1 --MS}}
\newcommand{\michael}[1]{\todo{#1 --MP}}
\newcommand{\martin}[1]{\todo{#1 --MK}}
\newcommand{\nima}[1]{\todo{#1 --NK}}
\newcommand{\pascalj}[1]{\todo{#1 --PJ}}

\newcommand{\chart}[1]{node~\textsc{#1}}

\ifdefined\notodocomments
  \renewcommand{\todo}[1]{\relax}
\fi

\newif\ifanonymous
\anonymoustrue

\newcommand{\anonurl}[1]{\ifanonymous URL removed for anonymity.\else\url{#1}\fi}
\newcommand{\footnoteanonurl}[1]{\footnote{\anonurl{#1}}}

\def\|#1|{\mathid{#1}}
\newcommand{\mathid}[1]{\ensuremath{\mathit{#1}}}
\def\<#1>{\codeid{#1}}
\protected\def\codeid#1{\ifmmode{\mbox{\smaller\ttfamily{#1}}}\else{\smaller\ttfamily #1\xspace}\fi}

\newlist{researchquestions}{enumerate}{1}
\setlist[researchquestions]{label*=\textbf{RQ\arabic*}}

\newcommand{\CalledMethodsBottom}{\<@Call\-ed\-Meth\-ods\-Bottom>\xspace}
\newcommand{\CalledMethods}{\<@Call\-ed\-Meth\-ods>\xspace}
\newcommand{\EnsuresCalledMethods}{\<@En\-sures\-Call\-ed\-Meth\-ods>\xspace}
\newcommand{\MustCall}{\codeid{@Must\-Call}\xspace}
\newcommand{\MustCallAlias}{\codeid{@Must\-Call\-Alias}\xspace}
\newcommand{\MustCallUnknown}{\codeid{@Must\-Call\-Unknown}\xspace}
\newcommand{\CreatesMustCallFor}{\<@Creates\-Must\-Call\-For>\xspace}
\newcommand{\ResetMustCall}{\CreatesMustCallFor}

\newcommand{\trule}[2]{\[\frac{#1}{#2}\]}
\newcommand{\truleinline}[2]{\ensuremath{#1\mathrel{\vdash}#2}}
\newcommand{\hastype}[1]{\mathbin{:}\trtext{#1}}
\newcommand{\trcode}[1]{\codeid{\smaller\smaller #1}}
\newcommand{\trtext}[1]{\mbox{\smaller\smaller #1}}
\newcommand{\trquoted}[1]{\trcode{"}#1\trcode{"}}



\newcommand{\numTypeSystems}{11\xspace} 
\newcommand{\numModifiedTypeSystems}{2\xspace} 
\newcommand{\numProjects}{12\xspace}
\newcommand{\numLOC}{88,680\xspace}
\newcommand{\numHumanAnnos}{803\xspace}
\newcommand{\percentInferred}{39\todo{check}\%\xspace}
\newcommand{\warningReductionPercent}{45\todo{check}\%\xspace}
\newcommand{\tsSpecificLoC}{61\todo{check}\xspace}


\hyphenation{type-state}        
\hyphenation{null-able}         


\setlength{\leftmargini}{.75\leftmargini}
\setlength{\leftmarginii}{.75\leftmarginii}
\setlength{\leftmarginiii}{.75\leftmarginiii}

\newcommand{\prefigcaption}{\vspace{-5pt}}
\newcommand{\posttablecaption}{\vspace{-5pt}}

\addtolength{\textfloatsep}{-.25\textfloatsep}
\addtolength{\dbltextfloatsep}{-.25\dbltextfloatsep}
\addtolength{\floatsep}{-.25\floatsep}
\addtolength{\dblfloatsep}{-.25\dblfloatsep}

\newcommand{\zph}{\phantom{0}}
\newcommand{\zzph}{\phantom{00}}

\newcommand{\ie}{i.e.,\xspace}
\newcommand{\eg}{e.g.,\xspace}

\newcommand{\nullaway}{NullAway\xspace}
\newcommand{\annotator}{Annotator\xspace}
\newcommand{\nullable}{\codeid{@Nullable}\xspace}
\newcommand{\nonnull}{\codeid{@NonNull}\xspace}
\newcommand{\code}[1]{\texttt{\small #1}}

\newcommand{\basicbaseline}{SinglePrompt\xspace}
\newcommand{\agentbaseline}{mini-SWE-agent\xspace}

\newcommand{\tightpara}[1]{\noindent{\emph{#1.}}}

\title{LLM-Based Repair of Static Nullability Errors}

\author{Nima Karimipour}
\orcid{0000-0002-2599-7770}
\affiliation{%
  \institution{University of California at Riverside}
  \department{Computer Science \& Engineering Dept}
  \city{Riverside}
  \country{USA}
}
\email{nima.karimipour@email.ucr.edu}

\author{Pascal Joos}
\orcid{0009-0008-4124-7639}
\affiliation{%
  \institution{CISPA Helmholtz Center for Information Security}
  \department{SoftwareLab}
  \city{Saarbrücken}
  \country{Germany}
}
\email{pascal.joos@outlook.de}

\author{Michael Pradel}
\orcid{0000-0003-1623-498X}
\affiliation{%
  \institution{CISPA Helmholtz Center for Information Security}
  \department{SoftwareLab}
  \city{Saarbrücken}
  \country{Germany}
}
\email{michael@binaervarianz.de}

\author{Martin Kellogg}
\orcid{0000-0002-3185-2340}
\affiliation{%
  \institution{New Jersey Institute of Technology}
  \department{Computer Science}
  \city{Newark}
  \country{USA}
}
\email{martin.kellogg@njit.edu}

\author{Manu Sridharan}
\orcid{0000-0001-7993-302X}
\affiliation{%
  \institution{University of California at Riverside}
  \department{Computer Science \& Engineering Dept}
  \city{Riverside}
  \country{USA}
}
\email{manu@cs.ucr.edu}

\begin{abstract}
Modern Java projects increasingly adopt static analysis tools that prevent null-pointer exceptions by treating nullness as a type property. However, integrating such tools into large, existing codebases remains a significant challenge. While annotation inference can eliminate many errors automatically, a subset of residual errors---typically a mix of real bugs and false positives---often persists and can only be resolved via code changes. Manually addressing these errors is tedious and error-prone.
Large language models (LLMs) offer a promising path toward automating these repairs, but na\"ively prompted LLMs often generate incorrect, contextually inappropriate edits.
We present \toolname, a system that integrates LLMs into a structured workflow for resolving the errors from a nullability checker. 
\toolname's decision process follows a flowchart derived from manual analysis of 200 real-world errors. It leverages static analysis to identify \emph{safe} and \emph{unsafe} usage regions of symbols, using error-free usage examples to contextualize model prompts. Patches are generated through an iterative interaction with the LLM that incorporates project-wide context and decision logic.
Our evaluation on 12 real-world Java projects shows that \toolname resolves 63\% of the \num{1119} nullability errors that remain after applying a state-of-the-art annotation inference technique.
Unlike two baselines (single-shot prompt and mini-SWE-agent), \toolname also largely preserves program semantics, with all unit tests passing in 10/12 projects after applying every edit proposed by \toolname,
and 98\% or more tests passing in the remaining two projects.
\looseness=-1
\end{abstract}



\maketitle

\section{Introduction}\label{sec:intro}

As null pointer exceptions (NPEs) are among the most common sources of run-time errors in Java, substantial research has focused on preventing null dereferences~\cite{PapiACPE2008,10.1145/1108792.1108798,Loginov2008,DBLP:conf/oopsla/MadhavanK11}.
In production environments, NPEs can lead to unexpected crashes and system downtime. Preventing NPEs as early as possible---ideally at compile time---is therefore essential for building reliable software. Static nullness checkers that treat nullability as a type property have emerged as effective tools for detecting potential NPEs before deployment. Examples include pluggable type checkers like Uber's \nullaway~\cite{nullaway}, Meta's Nullsafe~\cite{PianykhZL2022-overwrite}, and the Checker Framework's Nullness Checker~\cite{PapiACPE2008,10.1145/1985793.1985889}. These tools offer lightweight integration into existing development workflows, and with JSpecify standardization of a semantics for nullness annotations~\cite{jspecify,jspecifyAnnounce}, there is an increasing demand for these tools.

However, adopting such checkers in existing codebases is a significant challenge:
for precise and sound checking, the checkers require accurate nullability annotations.
To reduce this manual effort, recent work~\cite{Kellogg2023,nullawayannotator,nullgtn,dietrich_et_al:LIPIcs.ECOOP.2023.10} proposes \emph{inference of nullability annotations}, which takes a program without nullability annotations and adds annotations automatically.
However, annotations alone are often insufficient. A recent study~\cite{nullability-comparison} analyzes versions of programs before and after modules are enrolled into nullness type checkers. The study concludes that beyond adding annotations, developers often perform additional code logic changes as part of the enrollment process.
Yet, annotation inference tools only add annotations or suppressions, and do not modify the program logic. Hence, fully satisfying a static nullness checker requires further code changes.
Across our benchmark, we find that after applying a state-of-the-art annotation inference technique, 24.8\% of the initial nullability errors remain. Thus, fixing these residual nullability errors -- if done manually -- significantly hinders the adoption of nullness checkers.

This work presents \toolname, the first automated system for repairing residual type-based nullability errors that annotations alone cannot resolve. Our approach is designed to assist the adoption of static type-based nullness checkers. \toolname runs \emph{after} existing annotation inference tools, addressing the remaining errors that those tools leave unresolved. We implement and evaluate \toolname in the context of \nullaway~\cite{nullaway} and its inference engine~\cite{nullawayannotator}, showing that it complements state-of-the-art~\cite{nullability-comparison} inference techniques by handling hard-to-fix cases through automated code editing.
\toolname guides its repair process using the structure of nullability errors and program semantics. These traditional data sources are augmented with
a large language model (LLM), which \toolname relies on for semantic reasoning and patch generation.

While LLMs have shown strong capabilities for repair~\cite{Xia2024a,icse2025-RepairAgent,Cheng2025}, we observe that resolving residual nullness checker errors by na\"ively prompting an LLM or using general-purpose software engineering agents~\cite{Yang2024SWEagent} is insufficient in practice: the LLM often takes the path of least resistance to satisfy the static checker, without resolving the underlying problem.
As a result, LLM-generated patches are often ``plausible but incorrect''~\cite{QiLAR2015} in the parlance of automated program repair:
they repair the surface problem (the error from the static checker) but break something else in the process.
Nullability is a pervasive and context-sensitive issue: establishing the provenance of a nullable value and reasoning about its appropriate handling may require navigating across multiple classes. In large codebases, assembling the right context for the model is nontrivial.
To better understand this challenge, we conduct a preliminary study of 200 residual errors from \nullaway after running annotation inference and construct a flowchart that
encodes a human's reasoning process for addressing each error. \toolname uses this flowchart to
guide its queries to the LLM.

Our key insight for constructing effective LLM queries is to decompose the codebase into \emph{usage regions}, indexed by the symbols involved in a nullability error. Given these regions, the approach identifies \emph{safe usage patterns} of these symbols across the codebase and uses them to synthesize patches that are both contextually grounded and project-consistent. This novel strategy, of identifying safe usage patterns based on nullability error information, allows us to inject examples directly from the codebase into the LLM prompts, providing the necessary context to identify a suitable fix.
Our idea is motivated by the plastic surgery hypothesis~\cite{Barr2014PlasticSurgery}, which states that fixes can usually be constructed based on existing code-fragments in the same codebase.

Evaluating \toolname on twelve real-world, previously-unannotated Java projects shows that \toolname resolves 63\% of the \num{1119} residual nullability errors that remain after applying a state-of-the-art annotation inference technique.
Importantly, \toolname's patches largely preserve semantics, with all unit tests (held-out from \toolname) passing in 10/12 projects after applying every edit proposed by \toolname, and 98\% or more tests passing in the remaining two projects.
In contrast, two baselines (single-shot prompt and \agentbaseline~\cite{mini-SWE-agent}) remove more errors, 72\% and 78\%, but cause test failures in most (8/12) projects.
Further, a manual assessment of patch quality shows that \toolname's patches are more likely to be acceptable to developers without changes (61-81\% more likely)
and less likely to be completely unacceptable (23-33\% less likely).
\toolname is comparable to the agentic baseline in speed and cost:
1.2 minutes and \num{15000} tokens (\$0.05 as of current pricing) per error---and \emph{much} faster and cheaper than manual repair.
Overall, \toolname resolves fewer static nullability errors, but has a dramatically lower rate of ``plausible but incorrect'' patches than the baselines, making it much more palatable
for automated deployment without direct human supervision.

In summary, this paper's contributions are:
\begin{itemize}
  \item We conduct a study of 200 nullability errors that remain after annotation inference and distill a decision flowchart that guides patch generation.
  \item We introduce \toolname, an LLM-based and static analysis-based patching system that resolves residual nullability errors that cannot be addressed through annotations alone.
  \item We evaluate our approach on \num{1119} nullability errors from 12 real-world Java projects, showing that it resolves 63\% of the errors, while largely preserving correctness and offering mostly human-acceptable patches.
  \item We release the dataset, implementation, and detailed manual assessments as part of our artifact for reproducibility (Section~\ref{sec:data-availability}).
\end{itemize}

\section{Background: Java Nullability Checkers}\label{sec:background}

Type-based nullness checkers~\cite{PapiACPE2008,10.1145/1985793.1985889, nullaway, PianykhZL2022-overwrite} rely on \emph{type qualifiers}~\cite{FosterFA99} to refine the type system by expressing whether a reference type may include \code{null}. These qualifiers are written in Java using annotations:
\nullable indicates that a reference may be \code{null},
and \nonnull indicates that it must not be.
By requiring these qualifiers on parameter, return, and field types, these checkers are fully modular, i.e., each method can be checked independently, resulting in faster type checking.
Types without an explicit qualifier are assumed to be \nonnull by default, except for local variables whose qualifiers may be inferred automatically~\cite{PapiACPE2008}.  

Checkers typically enforce two key rules: (1) a \nullable expression must not be dereferenced without a preceding null check, and (2) a \nullable value must not be assigned to a location declared as \nonnull. These rules are enforced through a subtyping relation in which \nonnull~\code{T} is a subtype of \nullable~\code{T}. Assuming that object fields are properly initialized and all code has been type checked, these constraints collectively ensure that the program is free of NPEs.

\begin{figure}[t]
\begin{subfigure}{0.495\textwidth}
\begin{codebox}
\begin{lstlisting}[language=Java]
 public void bar(Foo param) { 

   param.deref(); 

 }
 bar(null);/*@ \label{li:bar-error} @*/// checker error
\end{lstlisting}
\end{codebox}
\caption{Nullability error}
\label{fig:nullability-error}
\end{subfigure}
\hfill
\begin{subfigure}{0.495\textwidth}
\begin{codebox}
\begin{lstlisting}[language=Java]
 public void bar($\color{green!50!black}\texttt{+@Nullable}$ Foo param) {
$\color{green!50!black}\texttt{+\quad if (param != null) \{}$
     param.deref();
$\color{green!50!black}\texttt{+\quad \}}$
 }
 bar(null);
\end{lstlisting}
\end{codebox}
\caption{Error resolved}
\label{fig:nullability-resolved}

\end{subfigure}
\caption{Example of a nullability error and its resolution.}
\label{fig:nullability-example}
\end{figure}

For example, consider \cref{fig:nullability-error}.
\noindent Method \code{bar(Foo)} is called with a \code{null} argument (line~\ref{li:bar-error}), which violates the nullness checker's rules, as the parameter \code{param} is \nonnull by default. The checker would report an error at the call site, indicating that a \nullable value is being passed to a method that expects a \nonnull parameter. To resolve this, the developer must either ensure that the argument is non-null or modify the method signature to accept \nullable values as done in \cref{fig:nullability-resolved}.
\noindent With the changes applied, no type rule is violated, and the nullness checker confirms that no potential null dereferences remain.

While our conceptual contributions could be applied to any Java nullability checker, 
we build \toolname on top of \nullaway~\cite{nullaway}, a nullability checker developed at Uber, due to its widespread adoption.
\nullaway has a companion annotation inference tool, \annotator~\cite{nullawayannotator}, which we use to generate an initial set of annotations for a target program, before invoking \toolname.
We call any nullability error that remains after the automated annotation inference a \emph{residual nullability error}, or for brevity simply \emph{nullability error}.

\section{Motivating Example}

\begin{figure}[t]
\centering
\begin{codebox}
\begin{subfigure}[]{0.50\textwidth}
\begin{lstlisting}[language=Java,frame=none,backgroundcolor=]
 package model;
 class User {
$\color{NavyBlue}\texttt{+\quad @Nullable}$ MapView mapView;
$\color{NavyBlue}\texttt{+\quad @Nullable}$ public MapView getMapView() {
     return this.mapView;
   }
 }

 package controller;
 class Controller {
   public void updateLocation(User user) {
$\color{diffremovebg}\texttt{-\quad\quad move(user.getMapView().getCenter());}$$~\color{darkred}\texttt{// Error}$ /*@ \label{li:map-error} @*/
$\color{green!50!black}\texttt{+\quad\quad Point2D point = user.getMapView() == null}\quad ?$
$\color{green!50!black}\texttt{+   \quad\quad\quad\quad new Point2D.Double(-1, -1)}\quad :$
$\color{green!50!black}\texttt{+   \quad\quad\quad\quad user.getMapView().getCenter();}$
$\color{green!50!black}\texttt{+\quad\quad move(point);}$
   }
 }
\end{lstlisting}
\end{subfigure}
\hfill
\begin{subfigure}[]{0.46\textwidth}
\begin{lstlisting}[language=Java, firstnumber=19,frame=none,backgroundcolor=]
 package ui.dashboard;
 class Dashboard {
   private User user = new User();
   public Point2D getCenter() {
     if (user.getMapView() == null) { /*@ \label{li:map-safe} @*/
         return new Point2D.Double(-1, -1);
     }
     return user.getMapView().getCenter();
   }
 }
\end{lstlisting}
\end{subfigure}
\end{codebox}
\caption{
Code example to be enrolled into a nullability checker with a residual nullability error (line 12).
Annotations in {\protect\color{NavyBlue} blue} are added by an annotation
inference tool like \annotator~\cite{nullawayannotator}, and are part of \toolname's input.
\toolname proposes the change that removes the {\protect\color{darkred} red} line and replaces it with the {\protect\color{green!50!black} green} lines (lines 13-16). Lines 23-26 show a safe usage pattern of \code{user.getMapView()}.
}
\label{fig:motivating-example}
\end{figure}

\Cref{fig:motivating-example} illustrates the challenges of resolving nullability errors and how our approach addresses these challenges. 
The goal is to enroll this code in static nullability checking with no remaining errors. The nullability checker reports an error that field \code{mapView} is assumed to be non-null by default but is not initialized in the constructor. To address this error, automated annotation inference adds \nullable annotations on the field and its getter, which resolves the original errors.
However, the getter annotation causes a new error at line~\ref{li:map-error} due to the dereference of \code{user.getMapView()}.
Since this dereference error cannot be resolved by adding annotations, automated annotation inference cannot repair it: deeper reasoning and non-trivial edits are required.

A null dereference error can be addressed in several ways, but selecting the best fix is often non-trivial. While inserting a null check is a common and desirable approach, determining the correct behavior for the \emph{then} branch---i.e., when the value is null---is particularly challenging. Synthesizing this behavior often requires domain-specific knowledge to ensure that the resulting patch matches the intended semantics of the program.
Na\"ively prompting an LLM to fix this error yields a patch that, when \code{user.getMapView()} is null, simply skips the call to \code{getCenter()} and returns.
However, this patch is not semantically aligned with how the developers handle such a null value in other parts of the codebase.

Instead, for the dereference of \code{user.getMapView()} on line~\ref{li:map-error}, \toolname identifies a usage pattern in the \code{Dashboard} class where the same method is accessed safely via an explicit null check at line~\ref{li:map-safe}. This safe usage serves as a guiding example, enabling \toolname to synthesize a patch that eliminates the dereference error while maintaining semantic fidelity. For this case, \toolname determines that inserting a null check around the dereference, and for the \emph{then} branch, materializing a default value, is the most appropriate behavior. Among several possible defaults, \toolname synthesizes a value that aligns with the program's conventions, in this case returning a \code{Point2D} object at coordinates \code{(-1, -1)}.

This fix is possible because \toolname can extract and index both positive (safe) and negative (unsafe) usage patterns for a given code element across the entire codebase. By tracking not only where a symbol is misused (e.g., dereferenced without a check) but also where it is handled safely, \toolname can retrieve usage-consistent examples to guide the model's synthesis. These usage regions are not constrained to a fixed context window and may span multiple files, classes, or modules. As a result, \toolname can surface semantically meaningful context even when relevant usages are far from the error site, helping the model generate fixes that are both context-aware and aligned with the project's conventions.

\section{Preliminary Study} \label{sec:study}

To better understand the resolution strategies required to resolve
residual nullability errors, we conduct a manual analysis of 200
such errors drawn from the three benchmarks that
show the least reduction in the number of errors after inference
in the original \annotator study~\cite{nullawayannotator}. We choose
these benchmarks because they contain a significant number of errors that cannot be resolved
through annotations alone. While we include these three benchmarks
(retrofit, conductor, and litiengine) in our evaluation, 
excluding them has no impact on our overall conclusions (\cref{sec:threats}).

\begin{figure*}[t]
\centering
\includegraphics[width=\textwidth]{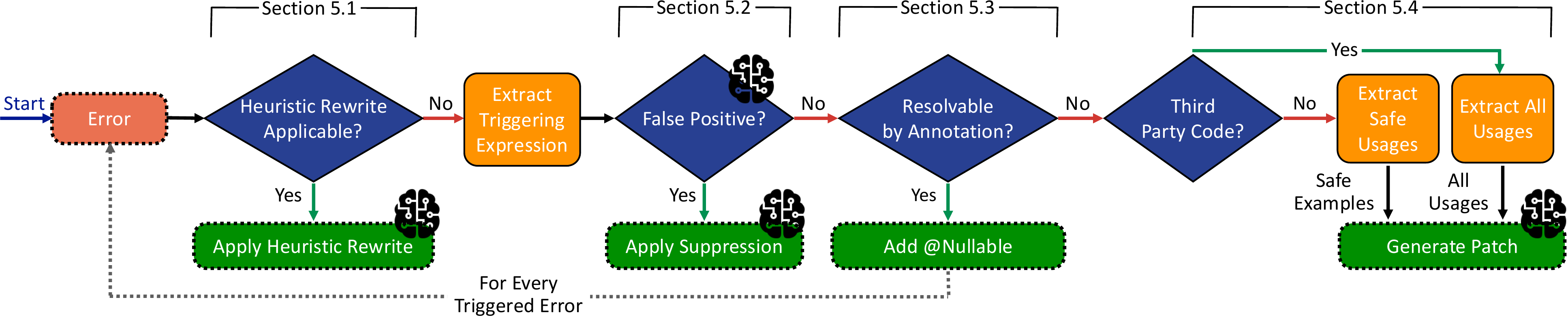}
\caption{Decision flowchart guiding \toolname's patch synthesis. Blue nodes denote decision points, orange nodes indicate action steps, green nodes are terminals where patches are generated.}
\label{fig:flowchart}
\end{figure*}

This preliminary study aims to uncover the types of information required to resolve each error, the decisions a human or automated tool must make, and the code transformations needed to produce a valid patch.  For each nullability error reported on the benchmarks after running annotation inference, we manually resolve it, tracing and documenting the following decisions:
\begin{itemize}
  \item The specific question that needs to be answered.
  \item The expected form of the answer.
  \item The resulting action required based on the answer.
\end{itemize}
We build a decision flowchart using this information incrementally. After analyzing each new error, we check whether our current flowchart accounts for each decision and action involved. If not, we expand the flowchart with new branches or refinements to reflect that error's resolution path. Over time, this iterative refinement converges to the decision flowchart shown in \cref{fig:flowchart}, which captures all the decision points and actions needed to address the variety in these 200 nullability errors.

This manual study is the empirical foundation for our approach: it grounds the flowchart in real-world error scenarios so that it can guide the repair process across diverse code patterns and error types. The next section describes how \toolname implements this strategy to resolve nullability errors through a series of automated decisions and transformations.

\section{Approach}\label{sec:approach}

Our goal is to automatically generate high-quality patches that eliminate nullability errors that remain after annotation inference, by imitating how a developer investigates and resolves such errors.
To structure this process, we use the flowchart-based decision protocol (\cref{fig:flowchart}) constructed as described in \cref{sec:study}.
Each error reported by the nullability checker has these components:
\begin{itemize}
  \item \textbf{Triggering expression:} the specific expression (e.g., local variable) whose nullability triggered the error.
  \item \textbf{Program point:} the code location (e.g., dereference site) where the error is reported.
  \item \textbf{Error type:} the category of nullability violation (e.g., dereference or return mismatch).
  \item \textbf{Error message:} the human-readable description of the error.
  \item \textbf{Suggested annotation (optional):} an annotation that would resolve the error, if there is one.\footnote{Some inference techniques, including \annotator~\cite{nullawayannotator}, heuristically omit annotations that trigger too many new errors.}
\end{itemize}

\noindent
The system launches an instance of \toolname for each error, each of which navigates the \cref{fig:flowchart} flowchart node by node. At each step, \toolname takes one of three actions: querying an LLM, running static analyses, or applying code transformations. This guided traversal continues until it reaches a terminal node and synthesizes a candidate fix.
This protocol-driven interaction is based on the intuition that incorporating surrounding code context and usage patterns can increase the likelihood that generated patches align with the program's intended behavior.
The following sections describe the flowchart in detail and explain how \toolname uses it to synthesize patches.

\subsection{Heuristic Code Rewriting}\label{sec:heuristic-rewriting}

First, \toolname checks whether a predefined heuristic null-safe rewrite pattern applies at the error site. Each heuristic consists of a precondition and a corresponding rewrite rule, designed to improve robustness against null dereferences while preserving (or minimally impacting) program semantics. \Cref{tab:rewrites} summarizes \toolname's rewrite patterns and their associated conditions.
If a matching rewrite pattern is applicable, the corresponding transformation is applied immediately. Otherwise, \toolname proceeds to the next decision node in the flowchart.

For example, when a receiver \code{x} of a call to \code{equals} is \nullable (precondition), \toolname replaces expression \code{x.equals(y)} with \code{Objects.equals(x,y)} (rewrite). This avoids a null dereference while preserving behavior when \code{x} is \nonnull. As another example, if a \nullable dereference \code{x.deref()} occurs in a method with a \nullable return type, and the nullness checker reports no errors at any call site due to the method returning \nullable (precondition), \toolname inserts an early \code{return null;} immediately after a null check on \code{x} (rewrite). Although this transformation alters the method's control flow, it respects the method's nullability contract and leverages existing null-handling at call sites to prevent a potential crash.

We perform heuristic code rewriting \emph{before} false positive suppression (\cref{sec:suppressing-false-positives}) for two reasons.  First, the transformations given in \Cref{tab:rewrites} are desirable regardless of whether the error is a false positive, as they provably eliminate the NPE risk, preserve semantics when the variable is not null, and behave appropriately when the variable is null.  For example, given a nullability error for \code{x.equals(y)}, rewriting to \code{Objects.equals(x, y)} preserves semantics when \code{x} is non-null and performs the desired comparison if \code{x} is null.  Second, our FP suppression leverages an LLM, and hence there is always some risk of inadvertently suppressing a true-positive report; applying the transformations of \Cref{tab:rewrites} has nearly no risk.

\begin{table}[t]
\footnotesize
\centering
\caption{Null-safe heuristic rewrites with preconditions that ensure semantic safety.}
\label{tab:rewrites}
\setlength{\tabcolsep}{2pt}
\begin{tabular}{
  @{}>{\raggedright\arraybackslash}p{2.5cm} 
     >{\raggedright\arraybackslash}p{4cm} 
     >{\raggedright\arraybackslash}p{7cm}@{}
}
\toprule
\textbf{Before} & \textbf{After} & \textbf{Precondition} \\
\midrule
\code{x.equals(y)} & \code{Objects.equals(x, y)} & \code{x} is \nullable \\
\code{x.hashCode()} & \code{Objects.hashCode(x)} & \code{x} is \nullable \\
\code{x.toString()} & \code{String.valueOf(x)} & \code{x} is \nullable \\
\code{x.deref()} & \code{if (x == null) return null; x.deref();} & 
\code{x} is \nullable, enclosing method is \nullable, safe call sites \\
\bottomrule
\end{tabular}
\end{table}

\subsection{Suppressing False Positives}\label{sec:suppressing-false-positives}

Our preliminary study shows several cases where the checker incorrectly flags an expression as being possibly null, leading to a false positive error. These misclassifications often arise due to limitations in the checker's reasoning---for example, failing to recognize guarding conditions, interprocedural guarantees, or method contracts that ensure non-nullness. In such cases, modifying the code to satisfy the checker may require intrusive or large-scale changes that are unlikely to be acceptable to developers. Therefore, we consider error suppressions (e.g., using casts or annotations) to be a practical and desirable strategy for resolving such errors. Notably, we observe that suppressions and casts are commonly used in existing enrolled codebases to silence similar errors, which confirms the findings of a recent study~\cite{fse2025_suppressions}. Consequently, \toolname attempts to detect such false positives and applies suppressions when appropriate.

Detecting false positives requires semantic understanding of the code and its context, difficult to capture with deterministic rules. We therefore prompt an LLM, which has shown to be effective at such detection tasks~\cite{Li2024withDOI,Li2024IRIS,Du2026}. 
\toolname extracts the method containing the reported error and constructs a specialized decision prompt to assess whether the triggering expression is guaranteed to be non-null at the error location. Each decision prompt follows a template that instructs the model to either (i) respond with a definitive answer when confident, or (ii) indicate uncertainty and request additional context. The model's reply is expected to follow a structured XML format. The full structure of these prompts is described in \cref{sec:llm-setup}.

Initially, \toolname includes only the method body in the prompt, as our preliminary study shows that this scope is often sufficient to identify false positives in practice, and eagerly including broader context significantly increases prompt size. 
If the model replies with uncertainty, it may request specific additional information, such as related method bodies, which \toolname retrieves and incorporates into a follow-up prompt.
If the model response indicates that the reported error is a false positive, \toolname inserts an explicit non-null cast at the error site, suppressing the error without modifying the program's behavior.

\begin{figure}[t]
\centering
\begin{minipage}{0.6\textwidth}
\begin{codebox}
\begin{lstlisting}[language=Java]
 @Nullable Session session;  
 void handle() {
   if(check()) {
$\color{diffremovebg}\texttt{-\quad\quad session.refresh(); }$
$\color{green!50!black}\texttt{+\quad\quad Nullability.castToNonnull(session).refresh(); }$
   }

 boolean check() { 
   return session != null; 
 }
\end{lstlisting}
\end{codebox}
\end{minipage}
\caption{Example of a false positive suppression.}
\label{fig:suppression-example}
\end{figure}

Consider the example in \cref{fig:suppression-example}. In this case, the model receives a prompt containing the body of \code{handle()}, where the nullable field \code{session} is dereferenced. The model replies with uncertainty and requests the implementation of \code{check()}. If the model infers that a null check within \code{check()} guarantees \code{session} is non-null at the dereference site, it classifies the error as a false positive, leading to the inserted cast.\footnote{See \url{https://github.com/uber/NullAway/wiki/Suppressing-Warnings\#downcasting} for more information on \nullaway{} casts.}

\subsection{Handling Triggered Errors}

If the error is determined to be a true positive, \toolname checks whether inserting an annotation can resolve it---which happens when the checker reports a \nullable value being assigned to a \nonnull location. In such cases, the fix is to add a \nullable annotation to the destination. After adding the annotation, \toolname reruns the type checker to identify new errors triggered by this change. Each newly triggered error is then fed back into the flowchart from the beginning and processed independently.
To avoid overlapping rewrites and ensure consistency across executions of the flowchart, \toolname maintains an up-to-date working copy of all previously rewritten regions. 
The final result is the union of all the changes computed across the original error and any subsequently triggered errors. Since handling triggering errors can, in principle, lead to an infinite sequence of new errors, \toolname has a budget for resolution attempts (see \cref{sec:llm-setup}).
This process differs from that followed by the previously run annotation inference tool, since annotation inference only adds annotations that reduce the total number of errors, while \toolname can address follow-on errors by applying further code changes.

\subsection{Safe and Unsafe Usage Regions} \label{sec:retrieving-usage-locations}

In our preliminary study, human developers first exhaust the previously-described techniques---heuristic rewrites, checking for false positives, and adding annotations---before resorting to larger code changes.
Following this strategy, only if these techniques fail, \toolname moves to the next decision node in the flowchart, where it synthesizes a code change by learning from usage patterns in the codebase.
The key idea in this step is leveraging the contrast between \emph{safe} and \emph{unsafe} usage patterns of the triggering expression, i.e., code locations where the same element is used without or with triggering a checker error, respectively.

To find these patterns, \toolname partitions the codebase into discrete \emph{regions}, defined as method bodies and field declarations, which serve as the semantic units within which symbols appear. We formalize this abstraction as a function: $\mathit{usageRegions} : \textit{Symbol} \rightarrow \{\textit{Method or Field Declaration}\}$.
This level of region granularity in practice preserves enough semantic information to guide the model with contextual clues about program intent. During repair, \toolname maintains an index that maps each symbol to its corresponding usage regions, and for each region, indicates whether the usage triggers a checker error due to usage of that symbol. 
Using this index, the system can efficiently retrieve representative \emph{safe} and \emph{unsafe} examples across the project and then provides all safe examples as part of the prompt for generating a patch.
In the rare case where no safe usages are found for a triggering expression, \toolname falls back to returning any usages.

To our knowledge, \toolname is the first repair approach to incorporate statically derived safe and unsafe usage contexts into a structured decision protocol for LLM-driven patch generation. This hybrid strategy increases contextual relevance while reducing overfitting to local code structure.

When retrieving usage examples to guide synthesis, \toolname distinguishes two cases based on the origin of the triggering expression. If the expression refers to a symbol declared in the codebase, \toolname directly retrieves safe and unsafe usages of that symbol across all regions. However, if the expression originates from third-party code, \toolname locates the closest receiver within the expression that \emph{is} declared in the codebase. It then collects all regions where this receiver is used, as these reveal the receiver's lifecycle and usage invariants. For instance, if the triggering expression is \code{map.get(key)} and \code{map} is a project-declared field, \toolname does not analyze all \code{Map.get()} usages globally. Instead, it focuses on how \emph{this specific} \code{map} instance is used. 

As an example, in line~\ref{li:map-error} of \cref{fig:motivating-example}, \code{user.getMapView()} causes a null dereference error that cannot be resolved by annotations. \toolname retrieves all usages of \code{getMapView()} across the project. It discovers that in the \code{Dashboard} class, the \code{mapView} field is checked for null before access (line~\ref{li:map-safe}) and a reasonable default is returned in the null case. This safe usage is included in the model prompt to guide synthesis. The model then produces a patch that mirrors this safe pattern by inserting a null check and similar default value for the line~\ref{li:map-error} dereference.

\subsection{LLM Setup}\label{sec:llm-setup}

Each \toolname query is stateless and constructed independently, relying only on the context accumulated up to that point. Queries fall into two categories: \emph{decision prompts} and \emph{patch prompts}.

\subsubsection{Decision Prompts}

Decision prompts assess whether a specific error is a false positive, e.g., whether a method cannot return \code{null}, to then traverse the flowchart. The model is expected to respond with either \code{Yes} or \code{No} to indicate agreement or disagreement, or \code{Unsure} when it cannot confidently answer. In the \code{Unsure} case, the model may request additional context---such as the method declaration and body, field declaration, or relevant usages---to make a more informed decision. Decision prompts are only used to decide if errors are false positives. The other decisions in the flowchart are decided based on static analyses, without involving the LLM.

\subsubsection{Patch Prompts}

Patch prompts request concrete edits to fix a specific error, given contextual evidence. These prompts typically include the triggering expression, safe usage examples extracted from the codebase, and the relevant enclosing method or field. The model is asked to return a rewritten version of the method that resolves the error. Each patch response is validated to ensure that the output is syntactically valid Java code. While this does not guarantee successful compilation, it ensures that the output can be parsed and further analyzed by downstream tools.

\subsubsection{Prompt Design and Validation}

To construct robust decision prompts, we begin with a labeled dataset of nullability errors drawn from the same benchmarks used in \cref{sec:study}. Each error is paired with a classification question and a ground truth answer obtained through manual inspection.
We first draft prompts and then use the LLM itself to refine them via meta-queries, similar to previous work that optimizes prompts using LLMs~\cite{Guo2023connecting}.
The meta-queries ask the model to suggest clearer phrasing, remove ambiguity, or better align with its reasoning patterns. This self-refinement loop helps make the final prompt templates interpretable and robust across contexts.

Each candidate prompt is tested on at least 20 examples from the benchmarks from \cref{sec:study}. To minimize overfitting, prompt tuning is limited to this subset. A prompt is accepted only if the model's responses consistently match the expected answer across all test cases. All prompt responses must follow a fixed XML schema, and are parsed and validated; if an output fails structural validation, a repair loop is triggered. This loop supplies the model with the malformed output and parse error, prompting it to retry, up to five times per query. Each error is allocated a fixed interaction budget: a maximum of 50 total LLM queries, with local retry thresholds. In practice, this budget ensures scalability while preserving sufficient reasoning depth for complex cases.

\begin{figure}[t]
\centering
\begin{subfigure}[t!]{0.55\textwidth}
\begin{promptbox}{Prompt Example (Shortened)}
\scriptsize\ttfamily
\#\#\# TARGET NULLABILITY ERROR \\
In line \textcolor{blue!60!black}{\texttt{move(user.getMapView().getCenter());}} the expression \textcolor{blue!60!black}{\texttt{getMapView()}} is nullable, causing a \textcolor{blue!60!black}{\texttt{null dereference}} error.\\
\textbf{Method:}\\
\textcolor{blue!60!black}{\texttt{public void updateLocation(User user) \{}} \\
\hspace*{1em}\textcolor{blue!60!black}{\texttt{move(user.getMapView().getCenter());}} \\
\textcolor{blue!60!black}{\texttt{\}}}\\

\#\#\# EXAMPLE \\
See a safe usage example of the \textcolor{blue!60!black}{\texttt{method getMapView()}}:\\
\textcolor{blue!60!black}{\texttt{public Point2D getCenter() \{}} \\
\hspace*{1em}\textcolor{blue!60!black}{\texttt{if (user.getMapView() == null) \{}} \\
\hspace*{2em}\textcolor{blue!60!black}{\texttt{return new Point2D.Double(-1, -1);}} \\
\hspace*{1em}\textcolor{blue!60!black}{\texttt{\}}} \\
\hspace*{1em}\textcolor{blue!60!black}{\texttt{return user.getMapView().getCenter();}} \\
\textcolor{blue!60!black}{\texttt{\}}}\\

\#\#\# TASK\\
Based on the example provided above, please rewrite the method to resolve the error.\\

\#\#\# GUIDELINE, RESPONSE FORMAT, OUTPUT EXAMPLE\\
\emph{(Omitted for brevity)}
\end{promptbox}
\end{subfigure}
\hfill
\begin{subfigure}[t!]{0.44\textwidth}
\begin{promptbox}{Model Response}
\scriptsize\ttfamily
\textless response\textgreater \\
\hspace*{1em}\textless success\textgreater \\
\hspace*{2em}True \\
\hspace*{1em}\textless /success\textgreater \\
\hspace*{1em}\textless code\textgreater \\
\hspace*{2em}\texttt{public void updateLocation(User user) \{}\\
\hspace*{3em}\texttt{Point2D point = user.getMapView() == null}\\
\hspace*{4em}\texttt{? new Point2D.Double(-1, -1)}\\
\hspace*{4em}\texttt{: user.getMapView().getCenter();}\\
\hspace*{3em}\texttt{move(point);}\\
\hspace*{2em}\texttt{\}}\\
\hspace*{1em}\textless /code\textgreater \\
\textless /response\textgreater
\end{promptbox}
\end{subfigure}
\caption{Example of a prompt for patch generation, with the model response. \textcolor{blue!60!black}{Blue text} captures program elements extracted from the codebase; the remaining text is part of the static prompt template.
}
\label{fig:prompt-example}
\end{figure}

\Cref{fig:prompt-example} shows an example of \toolname's prompting strategy. It illustrates a \emph{patch prompt}, where \toolname asks the model to generate a fix using contextual cues.

\section{Implementation}
\label{sec:implementation}

Our implementation is built on top of the \annotator tool~\cite{nullawayannotator}, which infers annotations to ease the adoption of nullability checkers. \annotator provides an extensible indexing framework over the input codebase, enabling efficient querying of program elements and their usage contexts.
We use the JavaParser~\cite{javaparser} library to parse Java source code and apply synthesized patches.

\paragraph{Region and Symbol Indexing.} 
\annotator constructs a global index that maps each triggered nullability error to the corresponding \emph{triggering expression}, the underlying \emph{symbol}, and the \emph{code regions} where that symbol appears. We extend this index to further classify regions into \emph{safe} and \emph{unsafe} usage regions, depending on whether they do or do not trigger errors (\cref{sec:retrieving-usage-locations}).

\section{Evaluation}\label{sec:evaluation}

\subsection{Experimental Setup}

\paragraph{Benchmarks}
In our evaluation, we use the same benchmarks used in the \annotator paper~\cite{nullawayannotator}. These benchmarks span diverse software categories and have not been previously checked with \nullaway, making them representative of the usage scenarios \toolname targets.  \Cref{tab:patch} reports the number of nullability errors remaining in each benchmark after applying annotation inference. These residual errors represent the lower bound of what inference alone cannot resolve, highlighting the remaining challenge. The table also includes the number of non-comment, non-whitespace lines of code for each benchmark to indicate the scale of each project.

\paragraph{Baselines}\label{sec:baseline}
We compare \toolname to two baselines, \basicbaseline and \agentbaseline. \emph{\basicbaseline} prompts the LLM once per error, where the prompt contains the nullability error along with the surrounding code.  We use 20 lines before and after the error location for surrounding code, or the entire enclosing method if it is smaller. \emph{\agentbaseline} follows an agentic approach~\cite{mini-SWE-agent, Yang2024SWEagent}, and is one of the top-performing SE-agents on the SWE-bench Verified benchmark~\cite{SWEbenchLeaderboard}.
The agent can freely explore the project to gather context, modify arbitrary files, and validate changes by running the build and \nullaway. We provide the agent with the target nullability error and instructions on how it can apply and validate changes. \agentbaseline uses shell commands to interact with the project and can thus perform arbitrary actions.

\paragraph{LLM Configuration}
We use \code{GPT-5.1} from OpenAI without reasoning for all experiments~\cite{openai}.
Both \agentbaseline and \toolname have the same maximum budget of 50 distinct prompts and \$0.50 per error.

\paragraph{Research Questions}
Ideally, we would expect a static nullness error repair tool to take a codebase with nullability errors and generate code patches that resolve all such errors with minimal code changes, in a way that aligns with the program's logic and structure, resulting in a compilable program with no remaining errors quickly and at a reasonable cost.  We evaluate two research questions to judge how well \toolname performs with respect to this ideal:
\begin{itemize}
    \item \textbf{RQ1:} How effective and high-quality are the patches produced by \toolname?
    \item \textbf{RQ2:} What are the end-to-end execution time and API cost of applying \toolname?
\end{itemize}

\begin{table}[t]
\centering
\footnotesize
\setlength{\tabcolsep}{1.6pt}
\caption{Comparison of \toolname and the baselines across two modes.
\textbf{KLoC} is the benchmark size in thousands of lines of code and \textbf{Errors} is the number of residual errors in the benchmark.
In the \textbf{Per Patch} section, \textbf{CE} is the number of generated patches with compilation errors, \textbf{NE} is the number of compiling patches that trigger at least one new nullability error, and \textbf{R} is the number of patches that fully resolve the target error (not triggering new nullability or compilation errors). In the \textbf{Combined} section, \textbf{R} is the total number of resolved target errors when selectively applying only patches that resolve their error without triggering any new unresolved errors.}
\label{tab:patch}
\begin{tabular}{l c c | ccc | ccc | ccc | c | c | c}
\toprule
\multirow{4}{*}{\textbf{Project}} & 
\multirow{4}{*}{\textbf{KLoC}} & 
\multirow{4}{*}{\textbf{Errors}} &

\multicolumn{9}{c|}{\textbf{Per Patch}} & 
\multicolumn{3}{c}{\textbf{Combined}} \\
\cmidrule(lr){4-12} \cmidrule(lr){13-15}

& & & 
\multicolumn{3}{c|}{\textbf{\toolname}} & 
\multicolumn{3}{c|}{\textbf{\agentbaseline}} & 
\multicolumn{3}{c|}{\textbf{\basicbaseline}} & 
\multicolumn{1}{c|}{\makecell{\textbf{\textsc{Null-}} \\ \textbf{\textsc{Repair}}}} & 
\multicolumn{1}{c|}{\makecell{\textbf{mini-} \\ \textbf{SWE-agent}}} & 
\multicolumn{1}{c}{\makecell{\textbf{Single-} \\ \textbf{Prompt}}} \\
& & &
\textbf{CE}  & \textbf{NE} & \textbf{R} &
\textbf{CE}  & \textbf{NE} & \textbf{R} &
\textbf{CE}  & \textbf{NE} & \textbf{R} &
\textbf{R} &
\textbf{R} & 
\textbf{R} \\
\midrule
conductor & 13.8 & 25   & 1  & 3 & 17 (68\%)   & 0  & 2 & 22 (88\%)     & 1  & 1  & 23 (92\%)    & 16 (64\%) & 23 (92\%) & 23 (92\%) \\
eureka & 9.1 & 24       & 0  & 3 & 21 (88\%)   & 1  & 2  & 21 (88\%)    & 1  & 4  & 18 (75\%)    & 21 (88\%) & 22 (92\%) & 19 (79\%) \\
glide & 27.3 & 87       & 7 & 15 & 60 (69\%)   & 6  & 11 & 65 (75\%)    & 2  & 14 & 68 (78\%)    & 63 (72\%) & 80 (92\%) & 69 (79\%) \\
gson & 8.6 & 32         & 0  & 6 & 17 (53\%)   & 1  &  1 & 28 (88\%)    & 1  & 1  & 26 (81\%)    & 18 (56\%) & 30 (94\%) & 25 (78\%) \\
jadx & 49.7 & 117       & 14 & 35 & 58 (50\%)  & 14 & 19 & 80 (68\%)    & 16 & 14 & 82 (70\%)    & 57 (49\%) & 86 (74\%) & 82 (70\%) \\
libgdx & 97.3 & 453     & 14 & 75 & 286 (63\%) & 23 & 71 & 343 (76\%)   & 56 & 71 & 293 (65\%)   & 301 (66\%) & 354 (78\%) & 310 (68\%) \\
litiengine & 32.3 & 152 & 17 & 34 & 95 (63\%)  & 5  & 23 & 117 (77\%)   & 17 & 12 & 116 (76\%)   & 86 (57\%) & 89 (59\%) & 119 (78\%) \\
mockito & 19.3 & 36     & 5  & 10 & 21 (58\%)  & 0  &  4 & 28 (78\%)    & 3  & 6  & 21 (58\%)    & 19 (53\%) & 26 (72\%) & 20 (56\%) \\
retrofit & 3.6 & 13     & 0  & 3  & 10 (77\%)  & 0  &  0 & 13 (100\%)   & 0  & 0  & 13 (100\%)   & 13 (100\%) & 12 (92\%) & 13 (100\%) \\
spring-boot & 39.9 & 84 & 2  & 18 & 61 (73\%)  & 6  & 13 & 60 (71\%)    & 11 & 17 & 53 (63\%)    & 60 (71\%) & 69 (82\%) & 54 (64\%) \\
wala-util & 19.7 & 73   & 2  & 16 & 31 (42\%)  & 6  &  6 & 57 (78\%)    & 7  & 10 & 52 (71\%)    & 31 (42\%) & 67 (92\%) & 56 (77\%) \\
zuul & 17.4 & 23        & 1  & 3  & 19 (83\%)  & 2  &  2 & 19 (83\%)    & 5  & 4  & 12 (52\%)    & 19 (83\%) & 20 (87\%) & 12 (52\%) \\
\midrule
\textbf{Total} & 338.0 & \num{1119}     & 63  & 221 & 696 (62\%)   & 64  & 154 & 853 (76\%)   & 120 & 154 & 777 (69\%)    & 704 (63\%) & 878 (78\%) & 802 (72\%) \\
\bottomrule
\end{tabular}
\end{table}

\begin{figure}[t]
\centering
\includegraphics[width=0.7\textwidth]{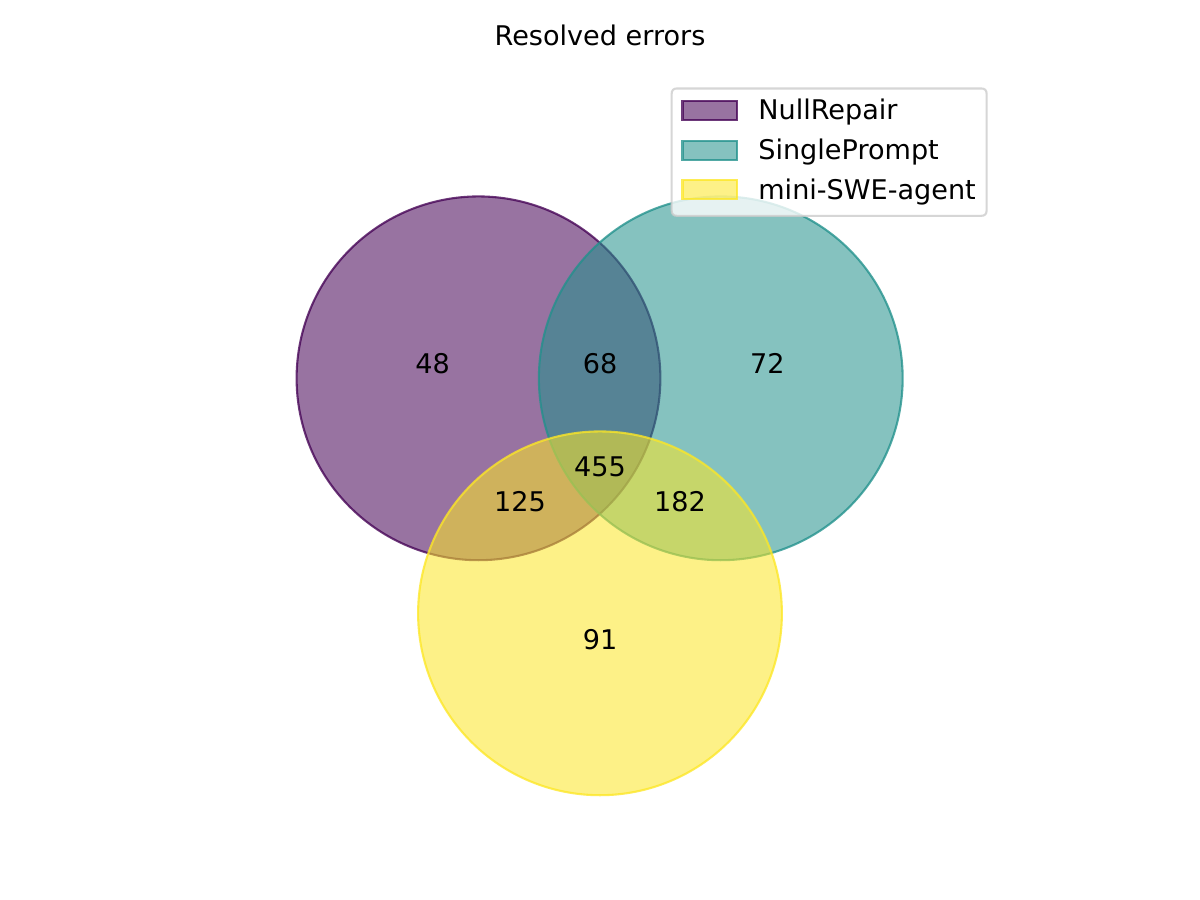}
\caption{Venn diagram showing the intersection of patches generated by \toolname, mini-SWE-agent, and SinglePrompt that fully resolve the target error (not triggering new nullability or compilation errors).}
\label{fig:venn-diagram-resolved}
\end{figure}

\subsection{RQ1: Effectiveness and Quality of Patches}\label{sec:effectiveness}

We aim to determine whether the patches generated by \toolname are acceptable to developers. Determining whether a fix is truly acceptable is a multidimensional question. There are often multiple valid ways to resolve a nullability error, depending on factors such as coding conventions, design intent, and local control flow structure. Rather than relying on a single fixed criterion to judge patch quality, we evaluate effectiveness through a series of criteria each capturing a different aspect of what makes patches useful and consistent with developer expectations.

\begin{itemize}
  \item \textbf{C1: Static Correctness:} Do patches compile successfully and resolve the targeted nullability error without introducing new unresolved errors?
  \item \textbf{C2: Dynamic Correctness:} Do patches preserve the original program behavior, as measured by the success rate of existing test suites?
  \item \textbf{C3: Manual Assessment:} Are patches logically aligned with the surrounding code and likely to be acceptable to a human developer in a code review?
\end{itemize}

We measure these criteria along two complementary axes: (1) a \emph{patch-level analysis} that evaluates whether each individual patch compiles and resolves its specific error, and (2) an \emph{aggregate-level analysis} that measures the total number of resolved nullability errors after selectively applying only the patches that both compile and fully resolve their targeted error. This simulates a realistic adoption scenario where only statically sound patches are retained.
For each residual nullability error, we attempt to generate a patch using both the baselines and \toolname. We then re-run \nullaway on the patched code to re-index any remaining or newly introduced errors.

\paragraph{\textbf{C1: Static Correctness}} \Cref{tab:patch}~(\textit{Per Patch}) presents the patch-level results, reporting the number of non-compiling patches each approach generates (\textit{CE}), the number of compiling patches that introduce new nullability errors (\textit{NE}), and those that fully resolve the original nullability error without introducing new errors (\textit{R}). 
We find that \toolname generates significantly fewer non-compiling patches compared to \basicbaseline. On average, for only 5.63\% of errors \toolname produces patches that fail to compile, compared to 10.72\% for \basicbaseline. This improvement is largely due to \toolname's use of contextual cues, such as method usage and inferred constraints, which guide patch generation toward semantically valid code. \agentbaseline has a similar compilation failure rate (5.72\%) to \toolname, as it can iteratively validate patches and refine them based on compilation feedback.
In terms of effectiveness, \toolname successfully resolves 62\% of errors on average (range: 42--88\%), \basicbaseline achieves 69\% (range: 52--100\%) and \agentbaseline reaches 76\% (range: 71--100\%), outperforming \toolname by 14 percentage points. 
\cref{fig:venn-diagram-resolved} shows the overlap between the sets of errors resolved by each approach. While \toolname resolves a significant portion of errors, there is a substantial subset that only the baselines resolve.
However, we observe that both baselines frequently rely on overly suppressive local edits that do not fix the underlying problem, which may obscure the actual quality of these fixes. The rate of newly introduced nullability errors also reflects this: \toolname triggers more (19.75\%) than the baselines (both 13.76\%). This trade-off suggests that \toolname often attempts more ambitious edits that engage deeper semantic issues, whereas the baselines prefer minimal, often suppressive changes.

\Cref{tab:patch}~(\textit{Combined}) presents the aggregate-level evaluation by measuring the total number of resolved nullability errors after selectively applying only those generated patches that resolve the respective target error. For this, \toolname{} iteratively processes errors and applies patches sequentially, only retaining those that reduce the total number of errors. Please note that these are not necessarily the same patches as generated for \textit{Per Patch}, as the codebase changes with each applied patch, such that subsequent patches are generated in a different context, potentially making the task easier or harder. 
In \textit{Combined} the baselines achieve a higher overall percentage of resolved errors (\basicbaseline 72\% and \agentbaseline 78\% in aggregate) compared to \toolname (63\% in aggregate). 

However, this advantage stems primarily from the baselines' aggressive use of guard conditions and explicit throw statements that silence errors without addressing their underlying cause.

\begin{figure}[t]
\centering
\begin{codebox}
\begin{lstlisting}[language=Java]
 @SuppressWarnings("unchecked")
 public <T extends Channel> T getChannel(ChannelDescriptor descriptor) {
   for (Channel array : arrays) {
     if (array.id == descriptor.id) return (T) array;
   }
$\color{diffremovebg}\texttt{-\quad return null;  // Error, as getChannel(ChannelDescriptor) is @NonNull}$ 
$\color{green!50!black}\texttt{+\quad throw new IllegalStateException("Channel not found for descriptor id: " + descriptor.id);}$
 }
\end{lstlisting}
\end{codebox}
\caption{Example of incorrect suppression by baselines, breaking program functionality at a call-site.}
\label{fig:baseline-incorrect-suppression}
\end{figure}

Consider the example in \cref{fig:baseline-incorrect-suppression}.
Both the \basicbaseline and \agentbaseline baselines forcefully silence a nullability error by explicitly throwing a run-time exception. While this suppresses the error from the checker, which correctly interprets the explicit throw as programmer intent, it breaks the code's functionality: 
A call-site \code{addChannel(...)} only creates a new channel after checking if \code{getChannel(ChannelDescriptor)} returns \code{null} and would therefore always trigger the run-time exception when trying to create a new channel.
\toolname runs out of queries attempting to resolve this error, and so does not propose a patch.
In a case like this, no patch is actually better than this (incorrect!) suppression.
However, in the error resolution numbers this example looks like a success for the baselines.

In other cases, we observe that the baselines often guess default values for fields that are inconsistent with the program's logic.
For example, in a class \code{TaskModel} a \code{status} of \code{null} can indicate an uninitialized state. The default value guessed by the baselines, \code{Status.SCHEDULED}, is wrong, as the \code{TaskModel} object is not necessarily scheduled at creation time, potentially masking errors and leading to unexpected program behavior.

\paragraph{Multi-file patches} In the per-patch evaluation, NullRepair creates 62 (5.5\%) multi-file patches that resolve the targeted error without introducing new errors or test failures (changing up to 8 different files).  
SinglePrompt produces zero multi-file patches, and mini-SWE-agent produces 12 (1\%) (with at most 3 files changed).

\begin{findingbox}
\textbf{C1:} \toolname patches compile and fully resolve the target errors in 62\% of cases. When combining patches, \toolname reduces total nullability errors by 63\% overall.
\end{findingbox}

\begin{table}[t]
\centering
\footnotesize
\caption{Comparison of \toolname and the baselines in regard to the number of failing unit tests per project after selectively applying patches that compile and resolve the target error without introducing new errors.}
\label{tab:tests}
\begin{tabular}{l c | ccc }
\toprule
\multirow{2}{*}{\textbf{Project}} & \multirow{2}{*}{\textbf{Tests}} & \multicolumn{3}{c}{\textbf{Number of Failing Unit Tests}} \\
\cmidrule(lr){3-5}
& & \textbf{\toolname} & \textbf{\agentbaseline} & \textbf{\basicbaseline} \\
\midrule
conductor & 436   & 0 (0.0\%) & 38 (8.7\%) & 27 (6.2\%) \\
eureka & 116       & 1 (0.9\%) & 5 (4.3\%) & 3 (2.6\%) \\
glide & \num{5335}  & 0 (0.0\%) & 6 (0.1\%) & 42 (0.8\%) \\
gson & \num{1058}   & 0 (0.0\%) & 3 (0.3\%) & 0 (0.0\%) \\
jadx & 751       & 0 (0.0\%) & 628 (83.6\%) & 647 (86.2\%) \\
libgdx & 98        & 0 (0.0\%) & 0 (0.0\%) & 0 (0.0\%) \\
litiengine & 508   & 0 (0.0\%) & 149 (29.3\%) & 77 (15.2\%) \\
mockito & \num{2089}     & 34 (1.6\%) & 0 (0.0\%) & \num{1136} (54.4\%) \\
retrofit & 349     & 0 (0.0\%) & 0 (0.0\%) & 0 (0.0\%) \\
spring-boot & \num{3920} &  0 (0.0\%) & 403 (10.3\%) & 842 (21.5\%) \\
wala-util & 1       & 0 (0.0\%) & 0 (0.0\%) & 0 (0.0\%) \\
zuul & 342        & 0 (0.0\%) & 1 (0.3\%) & 48 (14.0\%) \\
\midrule
\textbf{Total} & \num{15003}  & 35 (0.2\%) & \num{1233} (8.2\%) & \num{2822} (18.8\%) \\
\bottomrule
\end{tabular}
\end{table}

\paragraph{\textbf{C2: Dynamic Correctness}} To measure C2, we evaluate the patches by running each benchmark's unit tests after applying all generated patches. Similar to the aggregated setting in C1, we apply all selected patches from either \toolname or the baselines to the benchmark code and then execute the tests. The results, shown in \cref{tab:tests}, reveal a striking contrast: \toolname produces \textit{test-clean} patches (i.e., no test failures) in all but two benchmarks, whereas both baselines introduce at least one failing test in all but four benchmarks. Note that some of the benchmarks have extensive test coverage, with over 1,000 unit tests, and still \toolname preserves full correctness.

This result highlights a critical limitation of the baselines' patching strategies: while they may achieve a higher reduction in \nullaway errors, they often do so by aggressively silencing errors (e.g., via guard statements or broad null checks) without regard for preserving program behavior. These shallow fixes may satisfy the static checker, but they fail to maintain run-time semantics, leading to test failures. In contrast, \toolname leverages rich contextual signals, such as safe usage patterns, to synthesize semantically appropriate patches. The consistency of passing tests across a diverse set of benchmarks strongly supports the reliability and practical safety of our approach.

\begin{findingbox}
\textbf{C2:} \toolname produces \textit{test-clean} patches in all but two benchmarks, whereas both baselines introduce at least one failing test in all but four benchmarks.
\end{findingbox}

\paragraph{\textbf{C3: Manual Assessment}}To measure C3, we conduct a manual assessment guided by a scoring protocol. Importantly, we do not claim that a patch is definitively acceptable to the original developers of each benchmark---the benchmark projects are not actually adopting \nullaway at this point, so there is no incentive for their developers to dedicate time to review patches that fix \nullaway errors.
Instead, we define a structured rubric and scoring scheme that allows us to assess the \emph{likelihood} of acceptability. Patches are rated as \textit{1 - likely acceptable}, \textit{2 - needs work}, or \textit{3 - likely unacceptable}, based on whether they address the underlying issue correctly, preserve intended semantics, and follow idiomatic usage patterns.
\looseness=-1

We apply this evaluation to a total of 225 patches, consisting of 75 error cases for which both \toolname and the baselines produce fixes and remove the error. 
Each error is associated with three anonymized candidate patches---one from each approach---which are presented side by side in randomized order. All three reviewers are senior developers and authors of this paper, but are blinded to the source of each patch. Each patch is rated by at least two reviewers to ensure reliability and reduce individual bias. After initial independent scoring, reviewers discuss any disagreements to reach consensus.
The scoring criteria are:

\begin{itemize}
  \item Does the patch correctly identify the source of nullability and address it in a justifiable way?
  \item Are the values or behaviors introduced by the patch consistent with the rest of the program?
  \item Does the patch rely on suppression, and if so, is the suppression justified?
  \item Does the patch preserve object protocols and initialization semantics?
  \item Overall, would a human developer reasonably accept this patch in a code review?
\end{itemize}
Reviewers initially disagreed on 61 pairs. After discussing differences, the reviewers were able to reach consensus on all disagreements. The Cohen’s Kappa \cite{cohenskappa} value for their initial independent evaluations was 0.18, indicating low initial agreement. However, most of the disagreements (82\%) only differed by one point on the three-point scale, indicating that reviewers often had similar overall impressions but varied in their threshold for acceptability.

Our results show that \toolname produces 29 (39\%) patches rated as \textit{likely acceptable}, 16 (21\%) patches rated as \textit{needs work}, and 30 (40\%) patches that are \textit{likely unacceptable}. In total, 60\% of patches could be accepted after minor revisions during code review. The average score is 2.01. In comparison, the baselines \basicbaseline and \agentbaseline produce only 16 and 18 (21\% and 24\%) \textit{likely acceptable} patches, 14 and 18 (19\% and 24\%) patches that \textit{need work}, and 45 and 39 (60\% and 52\%) that are \textit{likely unacceptable}. This translates to average scores of 2.39 for \basicbaseline and 2.28 for \agentbaseline. Overall, \toolname outperforms both baselines in 28 cases, ties for the best patch with at least one baseline in 26 cases, and is outperformed by \textit{any} baseline in only 21 cases.
\looseness=-1

\begin{figure}[t]
\centering
\includegraphics[width=0.65\textwidth]{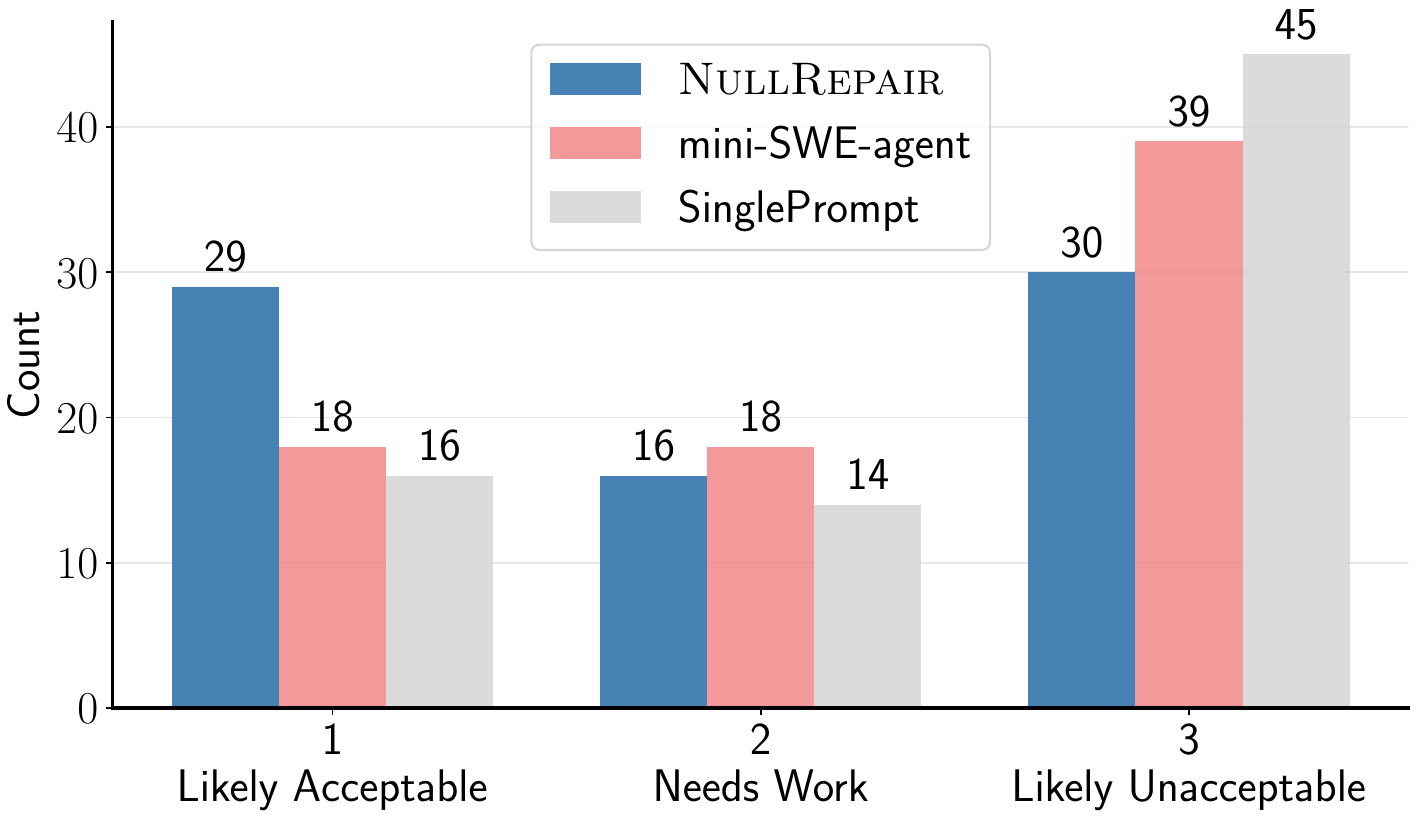}
\caption{The distribution of manual scores for patches generated by \toolname and the baselines.}
\label{fig:scores}
\end{figure}

A detailed score distribution (see \cref{fig:scores}) further illustrates these differences: while all approaches yield a similar number of patches rated as \textit{needs work}, \toolname produces more patches rated as \textit{likely acceptable} (61-81\% increase) and fewer rated as \textit{likely unacceptable} (23-33\% decrease) compared to the baselines.
Still, all three approaches generate a substantial number of patches that reviewers deem likely unacceptable, highlighting the inherent challenges of automated nullability repair. Relying only on the static correctness via resolved errors or dynamic correctness via existing test-suites is not sufficient to guarantee high quality fixes that are acceptable to developers.

When reviewers prefer patches produced by \toolname, it is due to their semantic precision and alignment with the program's intent. Rather than failing at the site of the error or setting incorrect default values, \toolname often applies coordinated changes across the codebase to safely propagate null-handling logic (7/28) or recognizes false positives that it then suppresses (21/28).
These patches reflect a deeper understanding of the surrounding logic and are generally perceived as more idiomatic and robust.
By contrast, both baselines mostly create very local edits, do not recognize false positives, or introduce hidden bugs.

\begin{figure}[t]
\centering
\begin{subfigure}[b]{0.66\textwidth}
\begin{codebox}
\begin{lstlisting}[language=Java]
$\color{diffremovebg}\texttt{- private Class<?> resourceClass;}$
$\color{green!50!black}\texttt{+ @Nullable private Class<?> resourceClass;}$
 ...
 List<Class<?>> getRegisteredResourceClasses() {
   return glideContext.getRegistry()
$\color{diffremovebg}\texttt{-\quad\quad\quad .getRegisteredResourceClasses(model.getClass(), }$
$\color{diffremovebg}\texttt{-\quad\quad\quad\quad\quad resourceClass, transcodeClass);}$
$\color{green!50!black}\texttt{+\quad\quad\quad .getRegisteredResourceClasses(model.getClass(),}$
$\color{green!50!black}\texttt{+\quad\quad\quad\quad\quad Nullability.castToNonnull(resourceClass), transcodeClass);}$
 }
 ... (further location where resourceClass is cast) ...
\end{lstlisting}
\end{codebox}
\caption{Likely acceptable patch by \toolname.}
\label{subfig:example-nullrepair-preferred-nullrepair-patch}
\end{subfigure}
\begin{subfigure}[b]{0.33\textwidth}
\begin{codebox}
\begin{lstlisting}[language=Java]
 void clear() {
$\color{diffremovebg}\texttt{-\quad resourceClass = null;}$
   ...
 }  
\end{lstlisting}
\end{codebox}
\caption{Likely unacceptable patch by the \basicbaseline baseline.}
\label{subfig:example-nullrepair-preferred-baseline-patch}
\end{subfigure}
\caption{Example where reviewers prefer \toolname's patch over both baselines.}
\label{fig:example-nullrepair-preferred}
\end{figure}

For example, consider the \toolname patch in \cref{subfig:example-nullrepair-preferred-nullrepair-patch}.
\toolname correctly identifies that the field \code{resourceClass} can be \code{null}, as a method \code{clear()} exists in the class that resets it to \code{null}. The approach therefore adds the \nullable annotation and handles multiple locations where \code{resourceClass} must not be null, by casting them to \nonnull (e.g., in the method \code{getRegisteredResourceClasses()}). This coordinated change preserves program semantics while resolving the nullability error.
In contrast, both baselines alter the behavior of the \code{clear()} method, by for example removing the line that sets \code{resourceClass} to \code{null} (\cref{subfig:example-nullrepair-preferred-baseline-patch}), effectively breaking the resetting behavior. This is an unintended change that can lead to incorrect downstream program behavior, as well as memory leaks.

However, in some cases reviewers favor one or both of the baseline approaches. Sometimes, \toolname creates overly complex fixes that reviewers cannot fully comprehend (3/21), or where they consider one of the baseline patches as more idiomatic (7/21). In 7 further cases, \toolname addresses the target error correctly but introduces an unrelated change or superfluous check. In the last 4 cases, \toolname makes an incorrect decision when determining whether the error is a false positive.
At least one of the baselines is preferred in each of these cases. By producing conservative and fail-fast fixes—such as raising exceptions or adding explicit null checks—the baselines ensure safety even without a full understanding of the broader context.

In ties, we observe three common scenarios. 
First, all approaches occasionally fail to recognize complex initialization or assignment patterns (e.g., a local variable that can be \code{null} in corner cases), leading to incomplete or unnecessary changes (12/26). Second, in 5 further cases, all three approaches fail to recognize that suboptimal annotations added by the annotation framework would need to be corrected.
Third, all three tools sometimes produce similarly effective local edits at the error site, requiring minimal context to resolve the issue acceptably (9/26).

\paragraph{Correctness of Classification} We also analyze the performance of \toolname's error classification step (\cref{sec:suppressing-false-positives}), on the same 75 error cases, by inferring the ground-truth classification from the reviewers' comments and disagreement-resolution discussions.
\toolname correctly classifies 27/33 true positive errors, and 33/42 false positive errors. This translates to a precision of 75\%, a recall of 82\%, and an F1 score of 78\%.
For some incorrect classifications, reviewers also expressed uncertainty about classification during their review, or initially disagreed with each other, indicating that these cases are inherently challenging even for human experts.

\begin{findingbox}
\textbf{C3:} \toolname produces higher-quality patches in 28 of 75 cases, ties for the best patch in 26 further cases, and 39\% of patches are likely acceptable to developers compared to baselines' 21\% and 24\%.
\end{findingbox}

Considering all three criteria together, \toolname can produce patches that are not only statically correct and test-clean, but also semantically aligned with the surrounding code. The baselines, while effective in some cases, often resort to suppressive strategies that compromise run-time behavior. This leads to a higher rate of test failures and lower overall patch quality.
\begin{findingbox}
\textbf{RQ1:} \toolname's patches are effective, with few test failures, and higher-quality than baseline patches, as judged by manual review.
\end{findingbox}

\subsection{RQ2: Cost and Performance}\label{sec:cost}

While the goal of \toolname is to produce developer-acceptable fixes, it must also do so efficiently and at reasonable cost. In this part of the evaluation, we examine the computational and resource-related cost of applying \toolname across the benchmarks.

We consider three performance metrics:
\begin{itemize}
    \item \textbf{Execution Time.} We report wall-clock time for \toolname and both baselines.
    \item \textbf{Number of Prompts.} We track the number of prompts sent to the LLM per error. For \agentbaseline this corresponds to the number of agent cycles.
    \item \textbf{Token Count and API Cost.} We report the number of tokens used per error to compute the API cost of running \toolname and the baselines.
\end{itemize}
Together, these metrics provide a holistic view of the performance and scalability of our approach.

\begin{table}[t]
\centering
\footnotesize
\caption{Comparison of \toolname and the baselines in regard to cost and performance. }
\label{tab:efficiency}
\begin{tabular}{l | cc | cc | cc }
\toprule
\textbf{Metric}
& \multicolumn{2}{c|}{\textbf{\toolname}}
& \multicolumn{2}{c|}{\textbf{\agentbaseline}}
& \multicolumn{2}{c}{\textbf{\basicbaseline}}
\\
\cmidrule(lr){2-3} \cmidrule(lr){4-5} \cmidrule(lr){6-7}
 &
\textbf{Per Project} &
\textbf{Per Error} &
\textbf{Per Project} &
\textbf{Per Error} &
\textbf{Per Project} &
\textbf{Per Error} \\
\midrule
\textbf{Execution Time (min)} & 113 & 1.2                  & 133 & 1.4    & 24 & 0.3 \\
\textbf{Number of Prompts} & 897 & 9.6            & 855 & 9.2     & 93 & 1.0 \\
\textbf{Total Tokens (k)} & \num{1411} & 15                  & \num{5632} & 60      & 79 & 0.8 \\
\textbf{Cost (USD)} & \$4.23 & \$0.05                  & \$4.05 & \$0.04    & \$0.32 & \$0.003 \\
\bottomrule
\end{tabular}
\end{table}

The results are presented in \cref{tab:efficiency}.
The total execution time of \toolname is 113 minutes, on average per project. However, this average is strongly impacted by the libgdx project, which contains many more nullability errors than the other projects. When excluding libgdx, the average execution time is roughly halved to 59 minutes. Per error the average execution time is 1.2 minutes. \agentbaseline is a bit slower with 1.4 minutes per error. Given that this is a one-time analysis, the execution time is reasonable for practical use.
\toolname creates on average 9.6 prompts per error. This is similar to \agentbaseline, which completes after 9.2 cycles per error. In contrast, the \basicbaseline baseline only sends one prompt per error.
\toolname has an average token usage of \num{15}\,k per error and \num{1411}\,k per project. Most of these tokens are uncached input tokens. Based on current API pricing, this translates to a cost of \$0.05 per error. We believe this is a reasonable overhead. \agentbaseline uses significantly more tokens (\num{60}\,k per error), but as the agent's prompts have large repetitive parts (agent history), most of these tokens are cached input tokens (\num{45}\,k per error), which are ten times cheaper. Therefore, the cost of \agentbaseline is similar to \toolname at \$0.04 per error. 

\begin{findingbox}
\textbf{RQ2:} \toolname is reasonably fast, averaging 113 minutes per project, and moderately priced, costing just \$0.05 per error.
\end{findingbox}

\section{Threats to Validity}{\label{sec:threats}}

The most important threat to the validity of our experiments is that three of twelve benchmarks in \cref{tab:patch,tab:tests} are used as the source
of the residual patches we study in \cref{sec:study} to design our flowchart. There is a risk that our technique will overfit to the residual errors in these three benchmarks, inflating our results. To mitigate this threat, we re-evaluate our experiments on just the nine benchmarks that are
not considered in our preliminary study, and find results that are nearly identical to the overall numbers in \cref{tab:patch,tab:tests}:
the per-patch repair rate (column R of Per Patch in \cref{tab:patch}) is 61.8\% on this subset of benchmarks vs. 62.2\% on all benchmarks; the combined repair rate (column R of Combined in \cref{tab:patch}) is 63.4\% on the subset, vs. 62.9\% on all benchmarks;
and the overall test failure rate is 0.26\% vs. 0.23\% (in \cref{tab:tests}).
None of these minor differences would impact our scientific conclusions in any way, so we are confident that the preliminary study's use of some
of the same benchmarks does not bias our overall results.

Another threat is generalizability: we evaluate our approach with only one LLM, one nullability checker, one annotation inference tool,
and on twelve projects in a single language. Other LLMs, checkers, inference tools, projects, or languages may require further innovation.
A further potential threat is data leakage. Although the LLM's training data may include examples of nullability errors, it is unlikely to have encountered the exact nullability errors present in our benchmark projects, or fixes addressing them, as, at the time of the model's training data cut-off date, only one of these projects has adopted \nullaway in their public repositories.
Another threat is our proxies for patch correctness: Tests can show the presence but not the absence of errors,
so passing tests are only weak evidence of correctness. However, tests are an improvement over only measuring error reduction, as much
of the prior work on annotation inference does.
Our manual evaluation of patch quality could also be biased. We mitigate this threat by
1) blinding ourselves to what approach patches stem from and randomizing the order in which each triplet of patches is presented,
2) having two authors independently review each patch, and 3) releasing the patches and judgments in our artifact, so other researchers
can check our reasoning.

\section{Related Work}

\paragraph{Automated Program Repair}
Researchers have proposed many techniques for repairing general software bugs~\cite{cacm2019-program-repair}, some of which are successfully deployed in practice~\cite{oopsla2019Getafix,Marginean2019,Williams2024}.
These approaches include heuristic, generate-and-validate approaches~\cite{LeGoues2012}, constraint-based search for semantically similar code~\cite{Ke2015}, and template-based repair~\cite{Liu2019a}.
Recent work builds on neural models, e.g., sequence-to-sequence models~\cite{Lutellier2020,Chen2021d,Ye2022a,Ye2024}, language models fine-tuned for the repair task~\cite{Jiang2023,Silva2024}, general-purpose LLMs~\cite{Xia2024a}, and via agentic LLM prompting~\cite{icse2025-RepairAgent,Cheng2025}.
Unlike our work, all these techniques repair run-time failures, typically visible via a failing test case.
\toolname{} cannot rely on a failing test as a validation oracle, and also must assess whether each error is a false positive.
Further work uses general-purpose agents for solving GitHub issues~\cite{Yang2024SWEagent, Zhang2024a}. We compare \toolname{} to \agentbaseline (\cref{sec:evaluation}), which is one of the leading openly available agents on SWE-bench Verified~\cite{Jimenez2023}.

\paragraph{Fixing Compilation Errors}
To help developers, especially novices, handle compilation errors (syntax errors, type errors, etc.), researchers have proposed
approaches that rely on sequence-to-sequence models trained for this purpose~\cite{Gupta2017} and LLMs~\cite{Joshi2023}, or combine a symbolic, error-correcting parser and a neural model~\cite{Sakkas2022}.
Unlike \toolname{}, these approaches assume that editing a single location is sufficient to fix an error.
Another key difference is that compiler errors must be fixed from the start, whereas we are looking at situations where one is adopting a type system later during the development process.

\paragraph{Fixing Static Analysis Errors}
Our work also relates to prior effort toward automatically fixing other kinds of static analysis errors.
One line of work learns from past fixes, e.g., by generalizing past fixes into a transformation rule~\cite{oopsla2019Getafix}, by expressing patches in a domain-specific language~\cite{Bavishi2019}, or by fine-tuning a sequence-to-sequence model~\cite{icse2024-PyTy,Berabi2024,Jin2023InferFix}.
Unlike these works, \toolname{} does not require past fixes or analysis-specific fine-tuning.
For instance, InferFix~\cite{Jin2023InferFix} targets potential null-dereference errors reported by the Infer static analyzer and relies on fine-tuning with a corpus of fixes for Infer warnings. In contrast, \toolname{} addresses arbitrary NullAway type system errors, which go beyond null-dereferences (e.g., assigning \nullable to \nonnull). It handles false positives, and employs safe usage regions from the codebase being repaired.
Similarly, NPEX~\cite{Lee2022} generates repairs based on null-handling patterns learned from a corpus, which is conceptually related to \toolname{}'s safe usage regions. However, NPEX is designed to repair null-dereference errors observed at runtime using stack traces as input, whereas \toolname{} addresses arbitrary statically reported NullAway errors. Consequently, \toolname{} can generate suppressions or annotation insertion repairs, which are important but out of scope for NPEX.

Another line of work defines code transformation rules by hand~\cite{Etemadi2023a}, which is inherently limited to a small set of rules and requires significant effort.
Finally, a third line of work deeply integrates a repair technique with a specific static analysis tool, e.g., by using an analysis that reasons about heap properties~\cite{Tonder2018}, internal feedback from a static analyzer~\cite{Zhang2025EffFix}, or symbolic execution to repair errors reported by a Datalog-based analysis~\cite{Liu2023}.
Unlike these deeply integrated approaches, \toolname{} is designed to work with any static analysis that reports nullability errors, such as \nullaway, in a black-box manner, and focuses on nullability errors rather than general memory errors.
Another difference between our work and the above is that we evaluate \toolname{} not only based on how many errors it removes, but also on its impact on the test pass rate, which we find to be more informative.

\paragraph{Other Related Work}
\toolname{} can be seen as another stepping stone to enable static nullability checking in existing code bases, which relates to prior work on type inference, especially for nullability annotation~\cite{Kellogg2023,nullawayannotator,nullgtn,dietrich_et_al:LIPIcs.ECOOP.2023.10,jha2024oppropberl}, but also for other kinds of type annotations~\cite{Hellendoorn2018,icse2019_NL2Type,fse2020_TypeWriter,Allamanis2020,Wei2020}.
Parts of our approach relate to prior efforts to automatically distinguish true from false positives in static analysis errors, which has been shown to be a difficult problem~\cite{Kang2022}, and can be addressed through a combination of static dependency analysis and LLM querying~\cite{Wen2024}.
The \textsc{Llm4sa} work~\cite{Wen2024} has not been applied to Java or nullability yet, but might further improve \toolname{}'s effectiveness at deciding when to suppress an error.
Suppressions of static analysis errors have been the subject of a recent study~\cite{fse2025_suppressions}, which shows that suppressions are relatively common in practice, confirming our design decision to support suppressions in \toolname{}.
MSA is a technique to port C code to Checked C, i.e., a memory-safe dialect of C~\cite{Mohammed2025}. 
Like our work, MSA also asks an LLM to add annotations or rewrite code to pass static checks that go beyond the core language, but for another language and another family of static checks.
\toolname{} relates to work on addressing run-time type errors in Python, with template-based~\cite{Oh2022} and learning-based approaches~\cite{Peng2024,hao24retyper}.
ReTyper~\cite{hao24retyper} also retrieves context from the current codebase to assist an LLM, but unlike \toolname{}, uses code similarity instead of static analysis to decide what context to retrieve.
Finally, \toolname's false positive decision-making allows the LLM to request additional context if needed. This relates to work~\cite{Loginov2008} on expanding the scope of static analysis on demand, based on whether the current context is sufficient for a conclusive decision.

\section{Conclusions}
\toolname advances the goal of fully automated migration to type-based nullability checkers by resolving nullability errors that remain after annotation inference. While our primary focus is on supporting from-scratch enrollment of existing codebases, \toolname is equally valuable in post-migration cleanup. For example, Uber has already used \annotator to migrate large codebases to \nullaway, but many hundreds or even thousands of error suppressions remain. \toolname can help eliminate these suppressions by synthesizing safe repairs.
By continuing the path laid down by annotation inference tools, \toolname brings us one step closer to the full automation of making software more reliable and robust against null-pointer exceptions.

\section*{Data Availability}\label{sec:data-availability}
\toolname is publicly available at \url{https://github.com/Pascal-Joos/NullRepair}, and archived with a DOI on Zenodo~\cite{NullRepairZenodo} to allow easy reproduction of our results.

\section*{Acknowledgements}
This research was supported in part by the National Science Foundation under
grants CCF-2223826, CCF-2312262, and CCF-2312263, the German Research Foundation (DFG; projects 492507603, 516334526, and 526259073), a gift
from Oracle Labs, and a Google Research Award.
The views expressed herein are those of the authors and do not necessarily reflect those
of our funders.

\bibliographystyle{ACM-Reference-Format}
\bibliography{local,plume-bib/bibstring-abbrev,plume-bib/types,plume-bib/dispatch,plume-bib/ernst,plume-bib/soft-eng,plume-bib/crossrefs,referencesMichael.bib}

\end{document}